# A complex social structure with fission-fusion properties can emerge from a simple foraging model

Gabriel Ramos-Fernández[1]*, Denis Boyer[2] and Vian P. Gómez[3]

[1]Centro Interdisciplinario de Investigación para el Desarrollo Integral Regional (CIIDIR) Instituto Politécnico Nacional - Unidad Oaxaca. A.P. 674 Oaxaca, Oaxaca 71230. México. Tel: +52 951 517 0400, Fax: +52 951 517 6000, E-mail: ramosfer@sas.upenn.edu

[2] Departamento de Sistemas Complejos, Instituto de Física, Universidad Nacional Autónoma de México, Apartado Postal 20-364, 01000 México DF, México.

[3] Instituto Latinoamericano de Comunicación Educativa (ILCE). Calle del Puente 45, Col. Ejidos de Huipulco, Delegación Tlalpan, C.P. 14380, México, D.F. México.

* Corresponding author

**ABSTRACT**

Precisely how ecological factors influence animal social structure is far from clear. We explore this question using an agent-based model inspired by the fission-fusion society of spider monkeys (*Ateles* spp). Our model introduces a realistic, complex foraging environment composed of many resource patches with size varying as an inverse power-law frequency distribution with exponent $\beta$. Foragers do not interact among them and start from random initial locations. They have either a complete or a partial knowledge of the environment and maximize the ratio between the size of the next visited patch and the distance traveled to it, ignoring previously visited patches. At intermediate values of $\beta$, when large patches are neither too scarce nor too abundant, foragers form groups (coincide at the same patch) with a similar size frequency distribution as the spider monkey's subgroups. Fission-fusion events create a network of associations that contains weak bonds among foragers that meet only rarely and strong bonds among those that repeat associations more frequently than would be expected by chance. The latter form sub-networks with the highest number of bonds and a high clustering coefficient at intermediate values of $\beta$. The weak bonds enable the whole social network to percolate. Some of our results are similar to those found in long-term field studies of spider monkeys and other fission-fusion species. We conclude that hypotheses about the ecological causes of fission-fusion and the origin of complex social structures should consider the heterogeneity and complexity of the environment in which social animals live.

**Keywords:** fission-fusion, spider monkeys, chimpanzees, agent-based models

## INTRODUCTION

Competition for food and predation risk are the most widely cited influences on the size and structure of animal groups (Alexander 1974; Bradbury and Vehrencamp 1976; Pulliam and Caraco 1984; van Schaik 1989). In primate societies, protection from alien male attacks (Wrangham, 1979), defense of group resources (Wrangham, 1980) and prevention of infanticide (Hrdy, 1977; rev. in van Schaik and Janson, 2000) also have been shown to be important determinants of group size and structure. However, when confronted with the wide variation in social structure existing among different taxa and even among populations of the same species, socioecological theory remains limited in its explanatory power (Janson 2000; DiFiore et al. in preparation).

Species with so called "fission-fusion" societies, such as chimpanzees (Goodall 1968), spider monkeys (Symington 1990) and dolphins (Connor et al. 2000), present both opportunities and challenges for socioecological theory. On the one hand, group size in these species changes over short temporal and spatial scales, such that large amounts of data can be gathered on a single population on the variation in group size and how it correlates with food abundance (e.g. Symington 1988; White and Wrangham 1988). On the other hand, the flexible nature of grouping patterns in fission-fusion societies creates methodological difficulties in defining, measuring and analyzing group size variation (Chapman et al. 1993), while the complexity of their foraging environments imposes difficulties in measuring resource abundance and distribution (Chapman et al. 1992).

In the studies carried out so far on fission-fusion primate species, no clear-cut pattern has emerged on the relationship between subgroup size and food availability. In a study on the interacting effects of the size, density and distribution of food patches upon the grouping behavior of spider monkeys and chimpanzees, Chapman et al. (1995) developed a simple, general model of how these three ecological variables should affect group size. They assumed that food patches could be found in one of three different configurations, each one leading to small or large subgroups: depleting and uniformly distributed, depleting and clumped and non-depleting patches. In their analysis, the authors found that only half or less of the variance in subgroup size in both spider monkeys and chimpanzees could be explained by habitat-wide measures of food abundance or variation in food patch size. Similarly, Newton-Fisher et al. (2000) found no correlation between subgroup size and habitat wide measures of food abundance; also, Anderson et al. (2002) found that party size in chimpanzees does not increase with food aggregation. Symington (1988) reported somewhat higher linear correlation indices for the average party size of spider monkeys and the size of feeding trees, although parties were larger at intermediate food patch densities than at low or high densities.

One reason for the lack of empirical support for socioecological explanations is that the development of testable, *a priori* predictions has lagged behind the accumulation of data and the formulation of *posthoc* explanations of why there is a correlation between, say, group size and the average size of feeding patches. This is especially true when considering that the real distribution and abundance of feeding patches found by forest-dwelling primates is far from being captured by idealized dichotomies such as clumped vs. uniform or large vs. small. Even when feeding for several days on only one species of fruit, it is likely that fruit-bearing trees of widely different size will be found, simply because of the age structure of the tree population. Recent studies (Enquist et al. 1999; Enquist and Niklas 2001) have found that tree size can be best described by an inverse power law frequency distribution, with similar exponent values across different forests throughout the world. In other words, small trees tend to be found in much higher numbers than large trees, but very large trees can sometimes be found. The importance of these “fat tails” in the size frequency distribution of feeding sources may be underestimated by averaging their size accross seasons or areas. The same argument applies to the size of animal groups, which has been found to vary, within a single species, according to power laws with “fat tails” (Bonabeau et al. 1999; Sjöberg et al. 2000; Lusseau et al. 2004).

What is required is a null model of social grouping that predicts the way in which subgroup size will vary when confronted with a realistic foraging environment. In such a model, agents would not interact through any social rules; rather, various agents may coincide at the same food patch, forming a group until they split as a consequence of the individual foraging trajectories. In a recent workshop on fission-fusion societies (Aureli et al. in preparation), DiFiore et al. (in preparation) proposed the use of agent-based models in which simple foragers and their emerging grouping patterns could be analyzed as a function of realistic environmental variation. This approach could allow behavioral ecologists to determine what would be the minimum conditions leading to variable grouping patterns and even non-random association patterns, simply as a consequence of the way in which animals forage in variable environments (DiFiore et al. in preparation).

In a spatially explicit model we developed recently (Boyer et al. in press), we showed that the complex foraging trajectories described by spider monkeys (Ramos-Fernández et al. 2004) could be the result of the distribution and abundance of food patches of varying size. In the model, a parameter defines the decay of the tree size frequency distribution and a single forager visits trees according to a least effort rule (minimizing the distance traveled and maximizing the size of the next patch). We found that complex foraging trajectories, similar in many aspects to those described by spider monkeys in the wild, emerged only at intermediate values of this parameter, that is, when large trees are neither too scarce nor to abundant (Boyer et al. in press). In the present paper we build on the same model, introducing several foragers into the same environment. We measure the tendency of these foragers to form groups and analyze their association patterns. Our purpose is not to test predictions of socioecological theory, but rather to develop a null model of the grouping and association patterns that should be expected to occur in a realistic foraging environment. We take advantage of the fact that this kind of model allows the manipulation of environmental variables, such as the relative abundance of feeding patches of different size, using only one parameter. We compare the results of the model with field data from spider monkeys.

## METHODS

### Model

We modelled the foraging environment as a two-dimensional square domain of area set to unity for convenience, and uniformly filled with 50,000 points (or targets) randomly distributed in space. These represent fruit-bearing trees. To each target *i* we assigned a random integer $k_i \geq 1$ representing its fruit content. All targets did not have the same fruit content a priori. At the beginning of the simulations, we set the fruit content of each tree to a random initial value $k_i^{(0)} \geq 1$, drawn from a normalized, inverse power-law probability distribution

$$p(k) = Ck^{-\beta}, \quad C = 1/\sum_{k=1}^{\infty} k^{-\beta} \qquad (1)$$

where β is a fixed exponent characterizing the environment, being the main parameter in the model. If β is close to 1, the range of sizes among the population is very broad, with targets of essentially all sizes. In contrast, when β >> 1, practically all targets have the same fruit content and the probability to find richer ones ($k_i^{(0)} = 2, 3\ldots$) is negligible.

This environment can be assumed to accurately represent a typical spider monkey habitat, where fruit content is known to be linearly dependent upon tree size (Chapman  et al. 1992; Stevenson et al. 1998), which in turn has been shown to vary according to an inverse power-law of the type of Eq. (1) in different tropical forests (Enquist et al. 1999). Exponent values measured in most forest types are in the range $1.5 < \beta < 4$ (Enquist and Niklas 2001, Niklas et al. 2003), while a typical spider monkey habitat in the Yucatan peninsula, Mexico, had a value of 2.6 (Boyer et al. in press). The number of trees was set according to the fruit tree densities in a typical spider monkey habitat (Ramos-Fernández and Ayala-Orozco 2003), which, depending on the species, lie between 3 and 300 trees per hectare (i.e. between 600 and 60,000 trees in a 200 hectare home range). The highest end of the range for the number of trees in a typical spider monkey habitat was chosen in order to obtain a wide range of variation in fruit content, similar to what monkeys would face when feeding on several species on a single day (Stevenson et al. 1998).

In this environment, we placed 100 foragers at different locations. These foragers represent spider monkeys or chimpanzees that forage for fruits among the existing trees. We chose 100 as it is close to what has been reported for spider monkey and chimpanzee community size (Goodall 1968; Symington 1990). Each forager was initially located at a randomly chosen target and moved according to the following rules: (a) the forager located at the tree number *i* next moved to a tree *j* such that the quantity $l_{ij} / k_j^{(0)}$ was minimal among all available tree $j \neq i$, where $l_{ij}$ is the distance separating the two trees and $k_j^{(0)}$ is the initial fruit content of tree *j*; (b) the forager did not choose a tree that it had already visited in the past. Thus, valuable trees (large *k*) could be chosen even if they were not the nearest to the foragers' position, as schematically illustrated in Fig. 1a. The ratio $l / k$ roughly represents a cost/gain ratio. Rule (b) was set according to the typical foraging trajectories of spider monkeys and other primates, who seldom retrace their own steps but rather visit a large number of distinct feeding sources before returning to a previously visited one (Milton 2000; Ramos-Fernández et al. 2004). In the model, time is discrete: during one time iteration (from *t* to *t*+1), a forager ate one unit of fruit of the tree it was located at. As several foragers could coincide at a given tree, at each iteration, the fruit content $k_i$ of a tree *i* decreased by 1 for each forager present on that tree. When the fruit content of the occupied tree reached zero, the forager(s) moved in one time unit to the next tree according to rules (a) and (b) above.

We used two different assumptions about the degree of knowledge that foragers had about the location and initial fruit content of trees. In the complete knowledge situation, foragers had perfect knowledge of the location of all trees and their initial fruit content, such that their choice, at every new move, was to visit the tree at which the ratio $l / k^{(0)}$ was minimum among all possible trees. In the partial knowledge situation, foragers only knew a random half of all possible trees (each forager knowing a different subset of trees). Thus, in the latter situation a forager could move in such a way that the ratio $l / k^{(0)}$ was not minimal among all the possible trees in the environment. Also, in both the complete and partial knowledge situations, due to the fact that a given forager only knew the initial size of targets not yet visited, it could visit targets that had already been depleted by other foragers (with a lower $k$ than expected). As explained above, when reaching an empty tree, the forager abandoned the tree in the next iteration. More details about the numerical procedures used to implement this model are presented in Boyer (2006).

Since each forager was unaware of the sequence of trees visited by others, a consequence of rule (b) above is that two foragers (A and B) meeting at a tree could split later on. This happened, for instance, when B had previously visited a target that A had not yet visited, but which A considered to be the next best target (Fig. 1b).

For each value of β and degree of forager knowledge, we ran a total of 50 different simulations in which trees and forager starting locations were randomly distributed in space. Each run consisted of 100 time iterations in which foragers either made a move to another tree or decreased the value $k$ of their current tree by 1.

**Analysis**

Given that our purpose was to evaluate subgroup formation by foragers and to compare this situation with what happens in real animals, we analyzed the resulting data sets in the same way as we would analyze field observations, particularly with regard to the following aspects:

*Subgroup size* was quantified by counting the number of times a forager was seen either alone or with different numbers of other foragers. The frequency distribution of subgroup size was obtained for different values of the resource parameter β and different degrees of forager knowledge, averaging over 50 independent runs and over all foragers. The average subgroup size refers to the average number of foragers with whom all 100 foragers were observed.

*Subgroup duration* was quantified by the average number of iterations that subgroups of a particular size lasted, averaged over 50 independent runs under various combinations of β and degree of forager knowledge.

*Relative affinity* was evaluated as the variance in the time each forager spent with each of the other foragers in the group. A high relative affinity implies that foragers were selective in their associations, limiting them mostly to a subset among all individuals they met, while a small relative affinity implies that all possible associations were more or less likely. For each forager *x*, we determined who it met (i.e. coincided at least once at the same tree) and for how long during the run. For all possible pairs, we computed an affinity $A_{x,y}$, defined as the amount of time units (not necessarily consecutive) that foragers *x* and *y* were together. For each forager *x*, we averaged $A_{x,y}$ and computed its variance over all the distinct *y*'s met by forager *x*. Dividing the variance of $A_{x,y}$ over its average, we obtained a non-dimensional number, lower than unity, that refers to the relative affinity of forager *x* with others: if close to 0, then *x* was "democratic" (i.e. it spent exactly the same amount of time with all foragers it met). If close to 1, forager *x* was "selective": it spent a lot of time with a few others, and a short time with most of the others it met. We then averaged this quantity over all independent runs and over all foragers, for a given combination of β and degree of forager knowledge. In order to compare this average relative affinity with what would be expected if encounters were at random, we obtained the same quantity for a randomized data set in which each forager *x* met the same number of distinct individuals *y*, and where the same total number of encounters made by *x* was distributed randomly among these *y*'s (for details on this randomization technique, see Whitehead 1999).

*Total bonds* refer to the number of distinct foragers met by a forager during a run. We obtained the average of this number, over all foragers and all independent runs, for various combinations of β and degree of forager knowledge.

*Strong bonds* refer to that subset of the total bonds that are more frequent than what would be expected from random and independent encounters. Therefore, it represents the number of "close associates" a forager had (Whitehead 1999). We determined, for a forager *x*, who it met during the run (foragers $y_1, y_2$...), and for how long ($A_{x,y_1}$, $A_{x,y_2}$...). Then we calculated $L_x$, the total number of meetings for forager *x* (the sum over all $A_{x,y_1}$, $A_{x,y_2}$). In parallel, we calculated the probability P($w$) that, among the total number $L_x$ of meetings, forager $x$ had $w$ meetings with the same individual if associations were at random. This was done analytically as follows: a number $L_x$ of bonds was drawn sequentially, from forager $x$ toward a randomly chosen forager included in its total bonds. Since $L_x$ and the total number of bonds are known from the simulation, we could compute P($w$) for these values. From this probability distribution we found the value $w_c$ such that $P(w > w_c) < 0.05$. The values $w > w_c$ are therefore very unlikely for random and independent meeting events. Strong bonds from forager $x$ to others were defined as those in which $A_{x,y} > w_c$. We obtained the average number of strong bonds over all independent runs, for various combinations of β and degree of forager knowledge.

*Weak bonds* refer to the total bonds that are not strong bonds.

*Clustering coefficients* for the networks formed by strongly bonded individuals refer to the probability that, if forager A has a strong bond with B and C, the latter are also strongly bonded among them (Newman 2000). Clustering measures the degree of transitivity in the social bonds of a network (or its degree of "cliquishness"). Let $r_x$ denote the number of strong bonds that forager $x$ has. Given the way in which we defined the strong bonds among foragers, the resulting network is not reciprocical *a priori*, but directed: a link going from $x$ to $y$, or out of $x$, does not imply that there is a link from $y$ to $x$; in other words, $y$ may be important for $x$, but $x$ may not be for $y$. The clustering coefficient $C_x$ is the ratio between the number of connections linking neighbors of x to each other and the maximum value, $r_x*(r_x-1)$, that this number can take (Newman 2000). Thus, a $C_x$ value of 0 means that any pair of foragers with which forager $x$ is strongly bonded are themselves not strongly bonded. Conversely, a $C_x$ value of 1 means that all the foragers strongly bonded to $x$ are also strongly bonded with each other. The clustering coefficient $C$ of the network was obtained by averaging $C_x$ over all foragers that had more than one strong bond and over the social networks obtained in the 50 independent runs, for each value of β and degree of forager knowledge.

*Relative size of the largest cluster* of a network refers to the number of individual foragers belonging to the largest cluster of the network divided by the total number of foragers. This is a measure of the cohesion of a network (Newman et al. 2002). A cluster is defined as an isolated part of the network, i.e. with no connections to other parts, that is itself not composed of various smaller isolated parts. Thus, any pair of nodes belonging to a cluster can be joined by at least one succession of bonds running through the cluster. Similarly, we define the *average cluster size* of a network as the number of individuals that do not belong to the largest cluster, divided by the number of clusters in the network (not counting the largest one). Both the relative size of the largest cluster and that of the average cluster were averaged for the 50 networks obtained in the independent runs, for each value of β and degree of forager knowledge. A network is said to *percolate* if the largest cluster contains a substantial fraction of the total number of nodes (see Newman et al. [2002] for a discussion in the context of social networks). When a network percolates, the size of the largest cluster (also called the giant cluster) is much larger than the average cluster size. We have performed the cluster analysis separately for the networks formed by the two types of bonds: i) total bonds, ii) strong bonds (see above).

It is important to note that, due to the high number of independent runs over which averages were calculated in each of the above analyses, standard errors were small (2-10% of the average value). Therefore, for clarity, results are shown without error bars.

## RESULTS

### Subgroup size

Figure 2a shows the normalized frequency distribution of subgroup size obtained in the model for various values of β and, for comparison, the values observed in a long-term study of two groups of spider monkeys (Ramos-Fernández and Ayala-Orozco 2003). Even though the majority of time foragers were alone, there is a clear effect of varying β upon the size of formed subgroups. Particularly for values of β between 2 and 4, the size of formed subgroups is sensibly larger than for the other values of β. When $\beta = 2.5$ and $\beta = 3$, the decay rate of the frequency distribution for subgroups in the model became indistinguishable from that of the real spider monkeys. Here, foragers could form subgroups of up to 17 individuals, although at a very low frequency. These values of β are close to the observed values in different forest types (Enquist and Niklas 2001), including one close to the study site where the data in Figure 2a come from, where a value of 2.6 was found (Boyer et al. in press).

Figure 2b shows the same data for the situation in which foragers had a partial knowledge of the location of feeding sites. As it can be seen, foragers formed smaller subgroups and the effect of varying β upon the size frequency distribution was less marked than in the situation with perfect knowledge.

The above can be seen more clearly when examining the way in which the average size of subgroups varied as a function of β, with full or partial knowledge of the location of feeding sites (Figure 2c). As can be observed, only in the full knowledge situation was there an increase in subgroup size at intermediate values of β, particularly at 2.5 and 3. That is, when foragers knew the location of all feeding sites, they formed the largest subgroups in an environment where large patches of food were neither too scarce nor too abundant compared to small patches.

**Subgroup duration**

Another way to analyze subgroup formation is by noting the time (in number of iterations) that associations lasted. As shown in Figure 3a, larger subgroups lasted less than smaller ones. For clarity, the graph shows subgroup size variation for only three values of β and the full knowledge situation. Subgroups of up to 3 foragers tend to last longer for β=2 than for other values of β. Focusing only on the most frequent type of association, Figure 3b shows the duration of subgroups of size 2 only, averaged over 50 independent runs as a function of β and for both knowledge situations. As β increased, associations were of shorter duration, although there was an intermediate range of values of β that had little effect on the average duration of pairs, particularly in the full knowledge situation. When foragers had only a partial knowledge of the location of feeding trees, pairs tended to last a shorter time, although this effect was more pronounced for values of β higher than 2. At β=2, large trees were relatively common and foragers stayed there for times that approximated half of the duration of the run, regardless of whether they had full or partial knowledge. Conversely, at β= 4.5, when there was a very small proportion of large feeding sites, foragers stayed a short amount of time at each one and visited a large number of different sites. In this situation, associations were of shorter duration.

**Preferential association**

In order to explore whether subgroups in the model were being formed by foragers at random, we calculated the relative affinity among foragers as the variance in the time they spent with different individuals. A high relative affinity implies that foragers were selective in their associations, limiting them mostly to a subset of all the individuals they met, whereas a small relative affinity implies that all the observed associations were more or less likely. We were interested in observing the effect of varying β upon the tendency to form preferential associations. However, the fact that foragers formed larger subgroups at particular values of β, implied that preferential associations could arise simply by chance. Thus, we calculated the expected relative affinities if associations occurred by chance, for each value of β.

Figure 4a shows the relative affinities expected randomly and those observed in the model, for different values of β, when foragers had full knowledge. At all values of β, relative affinities were higher than what would be expected if associations occurred by chance. The largest departures from random expectation occurred at intermediate values of β. Figure 4b shows the same data for the situation in which foragers had only partial knowledge of feeding sites. As before, relative affinities were higher than it would be expected by chance, but the difference is not so large as in the situation with perfect knowledge, particularly at high values of β.

**Network properties**

The relative affinities described above imply that, of all associations formed by a forager, some are more likely than would be expected by chance. In order to explore this skew in relative affinity in more detail, we calculated the total number of individuals met by each forager and, among these, determined who were the individuals that the forager met more often than would be expected purely by chance (strong bonds). Figure 5a shows the average number of bonds per forager as a function of β. As mentioned above, there was a clear effect of subgroup size upon the total number of bonds: there were more associations at intermediate values of β, particularly for β = 2.5 and 3, when the largest subgroups were formed (see Figure 2). Similarly, there was a clear effect of β upon the number of strong bonds, with the maximum number of strong bonds observed at β = 2.5. Figure 5b shows the same data for the partial knowledge situation. The effect of varying β was the same, upon the total number as well as the number of strong bonds.

Once we identified the strong bonds, it was possible to analyze the resulting social network and calculate the probability that if forager A had a strong bond with B and C, B and C also formed a strong bond between them (i.e. that there is transitivity in triadic relationships). This is the clustering coefficient of the social network (Newman 2000) and it varies from 0 to 1. Figure 5c shows the average clustering coefficients in the model as a function of β, for both knowledge situations. At low values of β, social networks had a high clustering coefficient in both the full and partial knowledge situations. However, as β increased, the clustering coefficients in the partial knowledge case fell sharply, while they remained high in the full knowledge case, up to β = 4.5, when they also decreased sharply.

**Percolation of the network**

Another structural aspect of the social networks that emerge in our model is the size of the largest cluster of linked foragers. If this cluster is much larger than the average cluster size (i.e. there is a “giant cluster”), a network is said to percolate. In a percolating social network, there is a high probability that any two individuals can be linked through other individuals that are themselves linked. The opposite of a percolating network is a fragmented one, in which there are many isolated clusters of individuals that never meet except amongst each other. Figure 5d shows the relative average size of the largest cluster formed by individuals who met at least once during the run (total bonds) or by only those individuals who met more often than expected by chance (strong bonds). A giant cluster is formed by the network of the total bonds at intermediate values of β. In the case of full knowledge and β = 2.5, the giant cluster contains about 20% of the foragers. The fact that these clusters are indeed the “giant clusters” is shown by the fact that the average size of the other clusters in the same network (data not shown) is much smaller, about 3.4 individuals. At both low and large values of β, no such percolation phenomenon is observed: the largest cluster size and the average cluster size are similar (2.8 and 1.1, respectively, for β = 4.5; 5.9 and 1.4 for β = 2.0). For the partial knowledge situation, despite the fact that it generates a smaller number of bonds per individual (Figure 5b), a giant cluster appears which is much larger: at β=2.5 it rises to 57% of the foragers. This suggests that the total bonds are formed in a more random way when the knowledge is limited, enabling easier connections between different parts of the network.

The network of the strong bonds exhibits fairly different properties than the network of total bonds at intermediate values of β. The clusters of strong bonds are smaller in size and no clear percolation property is observed at any value of β. The size of the largest cluster contains at most 7% of the foragers (β=2.5), a value not much larger than the average size of the other clusters in the same network (1.9 foragers). These values do not vary much with the degree of forager knowledge. These results indicate that individuals linked by strong bonds always form rather isolated structures. This property is consistent with the high values of the corresponding clustering coefficients (Figure 5c). If the total bonds are considered (which means adding all those bonds that are not strong, i.e. the weak bonds), the resulting network percolates at intermediate values of β, with clusters of strong bonds connected to each other via weak bonds. This situation is evident in Figure 6, which shows one of the networks that resulted at β=2.5 in a simulation with full knowledge. The weak bonds thus play an important role in the cohesion of the network when it is percolating.

## DISCUSSION

We have developed a simple foraging model that contains no algorithm specifying how foragers should interact. Our model focuses on the heterogeneity and structural complexity of the environment, summarized by the main parameter in the model, β. Despite its simplicity, the behavior generated by our model is quite rich (summarized in Table 1):

subgroups that vary their size in time are formed by foragers in response to the distribution and size of feeding targets; their size frequency distribution varies in response to β, being larger and more variable at intermediate values of this parameter, that is, when variation in tree size is intermediate, large targets being neither too scarce nor too abundant compared to small targets. Pairwise associations among foragers last longer at low values of β, when large targets are very common, but in these conditions the average size of subgroups is not the largest. In addition, there is little preferential association and few pairwise bonds that are more likely than random. It is at intermediate values of β that we observe the largest subgroups and where preferential associations arise. Foragers in these condition show many strong bonds and the social network formed by these strong bonds has a high clustering coefficient, a measure of the transitivity in the social bonds of the network (or the tendency of of foragers to form "clusters" or "cliques"). The weak bonds in that same network, on the other hand, connect different parts of the network, enabling it to percolate. At high values of β, when most targets are small, foragers group in smaller units with a short duration and their association patterns do not show as much preference as with other values of β. The social network in that situation does not percolate. Still, the foragers show a few strong bonds and the social network is moderately clustered at the local level.

Networks with properties similar to the ones described above have also been obtained in a model of mobile agents following stochastic trajectories and colliding with each other (González et al. 2006). In this study, though, the network structure does not arise from the complexity of the medium, which is uniform, but from particular kinetic rules for the agents.

In our model, foragers are able to decide which target to visit among several thousands of possible targets, representing the trees in a tropical forest that contain fruit at any given time. Even though a mental map of sorts can safely be assumed to exist in primate species (Janson 1998; Garber 2000), a full knowledge on the location and size of all possible targets is a strong assumption of our model. For this reason, we ran simulations in which foragers only knew a random half of the targets in the environment. The net effect of this "error" in the selection of the best target is that foragers form smaller subgroups, with less strong bonds and, consequently, a social network that is less clustered. However, even in the partial knowledge situation, there is a strong effect of intermediate values of $\beta$ upon the tendency of foragers to be in subgroups and to associate preferentially with others.

As stated in the Introduction, our purpose in developing this model was not to test existing hypotheses about how resources affect subgroup formation in fission-fusion societies, but to develop new predictions using numerical simulations, which can represent a complex environment better than simple conceptual models. The prevailing model on subgroup size and food resources in both chimpanzees and spider monkeys proposes that subgroups result from the interacting effects of the size and distribution of feeding patches (Symington 1988; Chapman et al. 1995). Large patches would feed more individuals than small patches, and the overall density of food patches would provide more opportunities for either a) traveling in large subgroups, as they would find food for all; b) dispersing in smaller subgroups as there would be no need to concentrate on a single patch. Depending on the assumptions made about predation pressure or other advantages of being in groups, the prediction on the effect of food density can be posed in both ways: larger or smaller subgroups in a high density of resources.

The study by Chapman et al. (1995) is an explicit test of these predictions. This study finds that a portion of the variance in subgroup size in spider monkeys (50%) and chimpanzees (30%) can indeed be explained by the overall density of food (the sum of the diameter at breast height or DBH of all available trees per hectare) and the distribution of food patches (variation in the number of fruiting trees per unit area). As density increases, subgroups tend to be larger. Also, when patches are farther apart from each other, subgroups tend to be smaller (Chapman et al. 1995). In another study, Newton-Fisher et al. (2000) found no correlation between subgroup size and food abundance in a chimpanzee group with a seemingly hyper abundant resource base. The authors of this study suggested that the relationship between food abundance and subgroup size is not linear, but curvilinear, such that "other factors" (Newton-Fisher et al. 2000, pp. 625) control the size of chimpanzee subgroups at high levels of food. In both studies, the authors attribute the weak correlations or the lack thereof to differences in how feeding competition affects age/sex classes (Chapman et al. 1995; Newton-Fisher et al. 2000).

Instead of developing *post-hoc* explanations, which eventually prevent the integration of social and ecological factors in the same model (Di Fiore et al. in preparation), it may be necessary to review the initial prediction of how food should affect grouping patterns. It is unlikely that, at any given time, spider monkeys or chimpanzees will find all patches to be small or to be widely spaced from each other. Most tropical tree species show clumped patterns in their distributions (Condit et al. 2000), and this pattern is highly dependent on scale, appearing uniform at small scales, clumped at intermediate scales and random (or Gaussian) at very large scales (Pélissier 1998). Also, the overall variation in tree size is best described by an inverse power-law (Enquist and Niklas 2001) and not by a Gaussian distribution. These important fluctuations imply that the mean may not be the best statistic to describe tree size. Moreover, both chimpanzees and spider monkeys may feed on several different species within a single day, let alone over periods of months or years (van Roosmalen and Klein 1987; Wrangham et al. 1996). Finally, the phenology of tropical trees is highly complex (Newstrom et al. 1994), with annual, sub-annual and supra-annual patterns all being relatively common (Bawa et al. 2003). These conditions result in a highly variable resource base, both temporally and spatially, which can hardly be captured by average temporal tendencies or overall spatial indices (Di Fiore et al. in preparation).

In our model, we use the variation in tree size as the independent variable, that is, tree size always varies but the parameter $\beta$ specifices exactly how this variation occurs. This parameter modifies the inverse power-law frequency distribution in Eq. (1). Tree-size distributions based on measurements of DBH are commonly characterized by exponents with values between 1.5 and 4 (Enquist and Niklas 2001), a range compatible with the values of $\beta$ that we considered in our model and with empirical measurements of $\beta$ in a typical spider monkey habitat (Boyer et al. in press).

In a previous version of our model (Boyer et al. in press), we explored the effect of tree size variation upon the movement trajectories of a single forager. We found that the longest and most variable movement trajectories, similar to those described by spider monkeys in the wild (Ramos-Fernández et al. 2004), appear at intermediate values of β. This situation is when the variance in the length of sojourns (or walks) given in the same direction is largest. This results from the foraging rule that the model introduces: when large trees are intermediate in their relative abundance, trajectories are composed of a series of short sojourns to visit mostly small trees, but every so often a large tree that is far away is worth the trip, so the forager takes a long sojourn to reach it. Conversely, when there are many large trees (small β) or when most are small (large β), the forager performs more regular trajectories composed of sojourns of similar length.

A similar pattern appears in the present version of the model in which the only change is the introduction of many foragers that move according to the same rules. It is only at intermediate values of β that foragers move in steps of variable size, often concentrating on small trees within a subregion but also traveling to large trees that are far away (data not shown). This explains why the largest subgroups are found at these values of β: foragers tend to consider rare, large trees as valuable and so they tend to coincide in them and, due to their size, to spend long periods of time in them. When β is small, foragers stay in the very common large trees, while at higher values of β, there are too few large trees and so foragers only spend small amounts of time in smaller trees that are close by. In both of these situations, they meet others rarely.

It is possible that, rather than the overall amount of food in the habitat of chimpanzees and spider monkeys, it is the relative importance of large trees when they neither too scarce nor too common that creates the conditions for large feeding aggregations to appear. Symington (1988) reported a nonlinear relationship (a second order polynomial) between patch density and the size of spider monkey feeding parties, which were larger at intermediate food patch densities. A similar result, but in another context, was obtained by Wilson and Richards (2000), who modelled a resource-consumer interaction in a spatially explicit environment. The authors found that, in the absence of rules by which consumers should interact, intermediate consumer densities (with a constant resource base) led to the formation of groups. The authors cite several other empirical examples where this occurs.

Our model simply presents the minimum conditions that could lead to a variable grouping pattern in a complex environment. It is clear that in real animals with fission-fusion societies, differences among age/sex classes in their reliance on food resources as well as their social strategies must play an important role in determining grouping and association patterns. However, upon close analysis of the composition of subgroups arising in the model, we found that, even when our model does not introduce any rule for their interaction or differences in their foraging strategies, foragers associate in nonrandom ways. For particular values of $\beta$, with full and partial knowledge, we find that foragers associate preferentially with certain others. This could simply be due to the fact that foragers are limited to particular regions of the environment, meeting only with those with whom, by chance, they share a common area. However, when taking only into account those individuals with whom an individual met at least once, there is still preference for some particular ones (Figures 4 and 5). Thus, we can conclude that this finding is not an artifact of the use of certain areas.

Preferential associations arise especially at intermediate values of β. The description of the foraging patterns can explain this: at low values of β, when there are many large trees, foragers only associate with those with whom they coincide upon reaching their first, common large tree. In a sense, this situation easily becomes "frozen," as foragers spend a large amount of time in each tree and there are many large trees in the environment. Conversely, at high values of β, associations last only short periods of time as they always occur in small trees. At intermediate values of β, when large trees are neither scarce nor common, foragers coincide with, and spend more time with, a larger subset of the available foragers. In addition, if this occurs at the beginning of the run, they may stay together for the whole run, as they would stay together throughout their subsequent foraging choices. At intermediate values of β, the fruit content of trees visited by a forager fluctuates widely (Boyer et al. in press), a fact that may explain why the time spent by the forager with other individuals (as measured by the affinity) also fluctuates so much. For these values of β, the foragers are also the most mobile, moving further away from their starting point (Boyer et al. in press). Therefore, it seems that the combination of two factors generates preferential association in our model: on the one hand, some heterogeneity in patch size, and on the other hand, relatively high forager mobility, allowing a large number of encounters.

The values of relative affinities we find in the model are comparable to those calculated from association matrices of two groups of spider monkeys by Ramos-Fernández (2001), using the same definition as in the present study. One group, with 9 adult individuals, had an average value of 0.21 ± 0.07 S.D. Another group, with 23 adults, had an average value of 0.59 ± 0.14 S.D. (Ramos-Fernández, unpublished data). Similarly, wild spider monkeys associate at detectable rates with the majority of the adults in their group (equivalent to the total bonds shown in Figure 5), but only 7-10 % of those associations are higher than it would be expected by chance (equivalent to the strong bonds in Figure 5; Ramos-Fernández 2001). Similar trends were found in chimpanzees by Pepper et al. (1999).

These results demonstrate that selective, nonrandom associations among animals (as defined by proximity) can arise simply from the way in which they forage and not necessarily as a result of their social relationships. We do not mean to imply that sex/age classes or social relationships are not important determinants of grouping patterns in social animals, but we find that nonrandom associations can emerge from the way in which foragers move in a complex environment. After all, social relationships in gregarious animals cannot have developed in an ecological vacuum: they must have developed within the existing grouping patterns that ecological conditions imposed.

A final aspect we explored was the structure of the social network formed by those foragers that were strongly bonded (i.e. those that associated more frequently than it would be expected by chance among all pairs that actually formed). This type of analysis of social networks has recently been applied to the social networks of dolphins, another species with a fission-fusion society (Lusseau 2003). One of the properties that defines the structure of a social network is its clustering coefficient, or the probability that if A is closely bonded with individuals B and C, the latter two are closely bonded too. This measure of the "cliquishness" of the social network formed by the foragers in our model is strikingly high. Social networks in wild spider monkeys have clustering coefficients between 0.26 and 0.30 (Ramos-Fernández, unpublished data), while the dolphin social network studied by Lusseau (2003) had a clustering coefficient of 0.303. In our model, the fact that clustering coefficients are close to 1 for most values of β, only in the full knowledge situation, may be a key to interpreting this result: when foragers coincide early in the run at a given tree, they will remain together for the remain of the run, which produces a large degree of selectivity and repeated associations among a few individuals. When foragers only know a random subset of all available trees, it is practically impossible that they will remain together for the whole run, as some trees will be known only by some but not all the foragers that may have coincided in a large tree at the beginning of a run.

Another property that characterizes the structure of a network is percolation, i.e. the possible existence of a "giant cluster" of individuals that can be linked through individuals that are themselves linked. The opposite of a percolating network is a fragmented one, in which there are many isolated clusters of individuals that never meet except amongst each other. The percolating properties of social networks of animals have received recent interest. The dolphin societies studied by Lusseau and Newman (2004) are formed of clustered sub-communities that are linked to each other by a few "broker" individuals. Two sub-communities were observed to interact very little while one of the brokers disappeared temporarily during the study (Lusseau and Newman 2004). These individuals are located at the periphery of the sub-communities but maintain the cohesion between them. Similarly, a typical social network emerging from our model includes relatively small clusters of strongly linked individuals. If the weak bonds are removed, the network formed by the strong bonds does not percolate. The network of the total bonds, however, does percolate at intermediate values of β, showing the importance of the weak bonds on its cohesion. In a different context, this so-called "strength of weak ties", has been long recognized to mediate interactions between agents belonging to different communities in human social networks (Granovetter 1973, 1983). In the case of animal fission-fusion societies, an intriguing aspect has been the fact that social relationships can be maintained in such a loose aggregation pattern (Kummer 1968; Smolker 2000; Ramos-Fernández 2005). While a percolating property based on a combination of strong and weak bonds has only been demonstrated in dolphins (Lusseau 2003), it remains to be determined whether the social networks of other species with fission-fusion societies also contain these structural properties. Our model points out at a mechanism by which these properties could emerge, simply out of the way in which animals forage in a complex environment.

Our model contrasts with that of te Boekhorst and Hogeweg (1994), who developed an agent-based model of a fission-fusion society in order to explain the differences in grouping tendencies between males and females. Even though the authors do not specify how trees in their model vary in size or how they are distributed in space, the model by te Boekhorst and Hogeweg (1994) contains rules by which foragers interact, that follow from the different behavioral strategies that both sexes should pursue, as proposed by Trivers (1972). As such, this model is not informative of the minimum conditions required for a variable grouping pattern to appear. Another modelling approach, aimed at understanding the emergence of social structure, has been taken by Hemelrijk (2000). She has modelled the emergence of dominance relationships as a consequence of the spatial distribution of individuals. Her models also incoporate rules by which individuals form groups, interact and modify their future social behavior according to these interactions. Both of the above examples of agent-based models are aimed at understanding the emergence of particular social relationships and structure. Thus, they incorporate differences among agents and rules by which they interact. Our model, in turn, does not make any assumption about the tendency to form groups or search each other. Rather, it is a spatially explicit depiction of agents foraging in a complex environment, as a result of which they form subgroups. As such, the results of our model should be used as a starting point to make more ellaborated predictions about the relationships we should find between subgroups and their environment in fission-fusion societies.

Our results lead us to propose the following predictions for field studies of fission-fusion social systems:

1) The relative abundances of small vs. large food patches should be better predictors of subgroup size than average food patch size, average food density or degree of clumpness.
2) Large patches may induce large subgroups that last for long periods of time, but due to the relative importance of large patches, an intermediate level of variation in patch size could induce the largest subgroups (albeit with a shorter duration). Therefore, we should observe large subgroups forming at large and infrequent patches and not in large and common ones.
3) Long trajectories could result from the relative importance of large patches. Therefore, we should observe these types of trajectories more frequently when food is found in less dense but very large patches. The resulting high mobility of foragers should enhance the frequency of encounters.

4) The social networks of fission-fusion species should be composed of several clusters of closely associated individuals that, in turn, are linked by looser relationships that nevertheless allow most individuals to remain within a single social network.

In conclusion, we have explored the minimum conditions that could lead to complex grouping and association patterns using an agent-based model that includes a spatially explicit representation of environmental variation. An intermediate degree of variation in the size of feeding patches can lead to larger feeding aggregations and more opportunities for social interactions to develop among foragers. Studies on the evolution of animal social relationships in complex environments must take these constraints into consideration.

**ACKNOWLEDGEMENTS**

Louise Barret, Colin A. Chapman, Anthony Di Fiore, S. Peter Henzi, Phyllis Lee and Julia Lehmann provided useful comments on a previous version of this manuscript, as did other participants at the workshop on Fission-Fusion Societies and Cognitive Evolution organized by Filippo Aureli, Colleen Schaffner and Cristophe Boesch and sponsored by the Wenner-Gren Foundation for Anthropological Research. We thank David Lusseau and an anonymous reviewer for fruitful suggestions during the review process. Funding was received from the following institutions: the Wenner-Gren Foundation, Tomás Brody visiting scholarship from the Institute of Physics, CONACYT (Grant number 40867-F), the National Autonomous University of Mexico (UNAM), the National Polytechnic Institute of Mexico (IPN) and the Fondo Sectorial CONACYT-SEMARNAT (project 0536). All experiments comply with the current laws of Mexico.

## FIGURE LEGENDS

Figure 1. (a) Trajectory map for a single forager. The size of targets represents their *k* value or fruit content. A forager starting at the target on the far right will go directly to the largest target, ignoring other smaller targets that were at shorter distances. (b) Trajectory map for several foragers. An additional forager to the one shown in Figure 1a (dotted lines), which started at the target on the far left would meet the first forager at the largest target (thus producing a fusion) and would stay with it, visiting the same targets until their history of previous visits would split them apart: the first forager would visit the target where the second forager departed, but the second would not visit this same target twice.

Figure 2. (a) Frequency distribution of subgroups of different size, for different values of $\beta$ and under the full knowledge situation. Each point corresponds to the average subgroup size in which all 100 foragers were found, averaged over all 50 independent runs. (b) The same as above, for the partial knowledge situation. For comparison, both (a) and (b) show data from two groups of spider monkeys (Ramos-Fernández and Ayala-Orozco 2003). (c) Average subgroup size as a function of $\beta$. The graph shows the average values for each of the distributions shown in (a) and (b). Standard errors are below 10% of the average values (not shown).

Figure 3. (a) Duration, in number of iterations, of subgroups of different size for three different values of $\beta$ and the full knowledge situation. (b) Subgroup duration as a function of $\beta$ and the degree of forager knowledge. In both figures, each point represents the average

number of iterations that all formed forager subgroups lasted in all 50 independent runs for each condition. Standard errors are below 10% of the average values (not shown).

Figure 4. Relative affinity in associations among foragers in the model. A value close to 1 shows a high skew toward particular individuals among all possible foragers met, while a value close to 0 implies an equal preference for all. Each value represents the average over all 100 individuals and over all 50 independent runs for each value of β. Shown is the same value of relative affinity for a randomized data set. See methods for the definitions. (a) Full knowledge situation; (b) partial knowledge situation. Standard errors are below 10% of the average values (not shown).

Figure 5. Average number of total bonds and number of bonds that can be considered as strong, i.e. much more common than expected by chance. Shown is the average number of bonds of each type over all 100 individuals and over all 50 independent runs in each condition. See methods for the definition of strong bond. (a) Full knowledge situation; (b) partial knowledge situation; (c) clustering coefficient calculated from the resulting social networks as a function of β and degree of forager knowledge. The coefficient is a measure of the "cliquishness" of the resulting networks, or the probability that if there is a strong bond between a forager A and foragers B and C, then B and C are strongly bonded between them too. Shown are the average coefficients for 50 independent social networks obtained in each condition. (d) Average size of the largest cluster in the social network formed by foragers who met at least once during the run (total bonds) or by foragers who met at higher rates than random expectation (strong bonds), under conditions of full or limited

knowledge, as a function of $\beta$. Each point represents the average of 50 independent runs for each value of $\beta$ or knowledge condition. Standard errors are below 10% of the average values (not shown).

Figure 6. Graphic depiction of one of the social networks that emerges in a situation with complete knowledge and $\beta = 2.5$ (not all foragers are represented). Black arrows correspond to strong bonds ($A \rightarrow B$ means that B is a strong associate for A), while grey lines correspond to weak bonds (see Methods for definitions). The figure clearly shows that the majority of foragers associate in clusters of strong bonds that are part of much larger clusters held together by weak bonds. The graph was obtained using the Pajek software (Batagelj and Mrvar 1998).

Table 1. Summary of main results. Subgroup size, duration of associations, relative affinity, number of strong bonds, cliquishness (clustering coefficients) and percolation of the network as a function of environmental heterogeneity (exponent $\beta$) and degree of forager knowledge about the location and size of trees in the environment.

Figure 1a

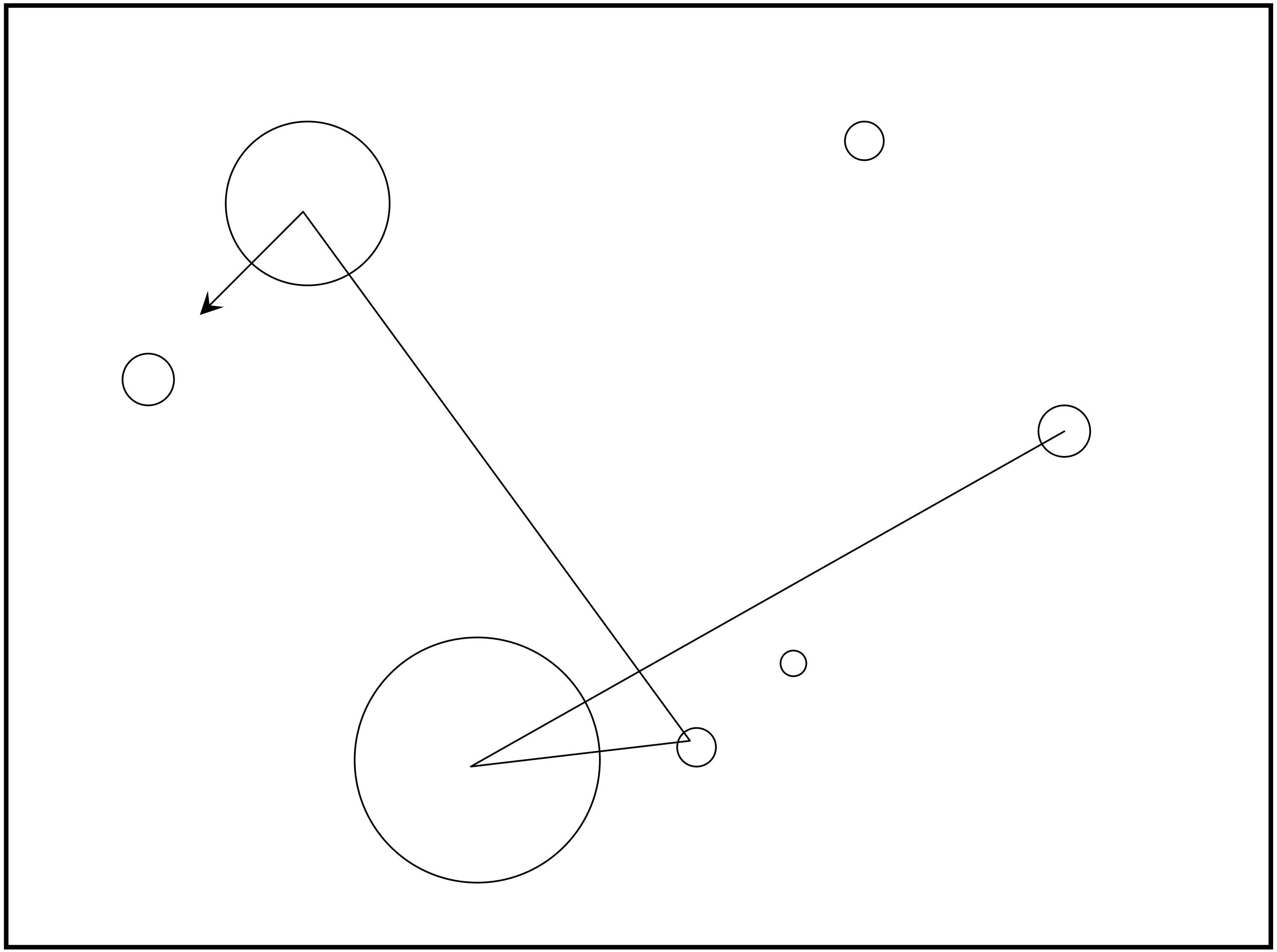

Figure 1b

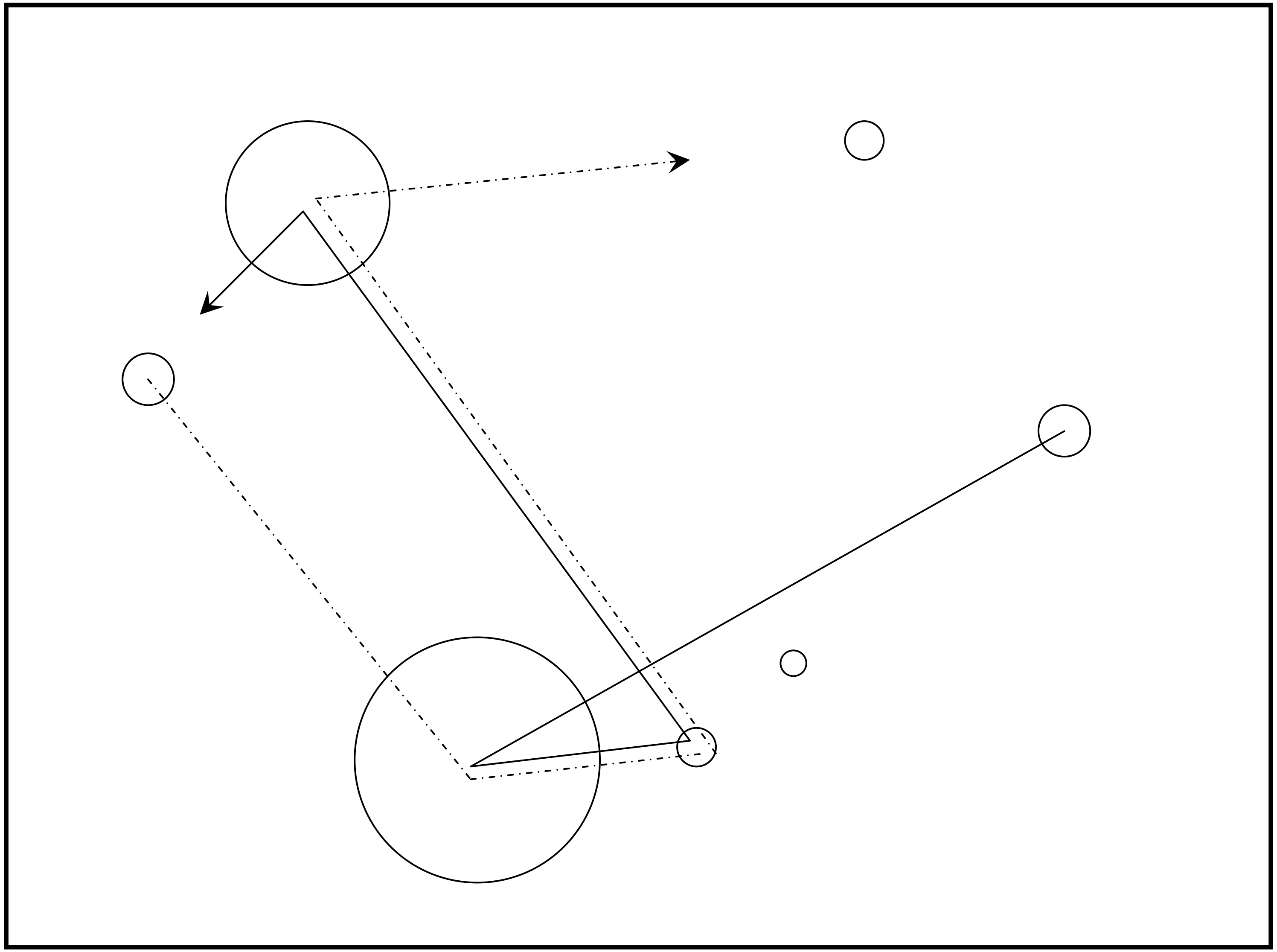

Figure 2a

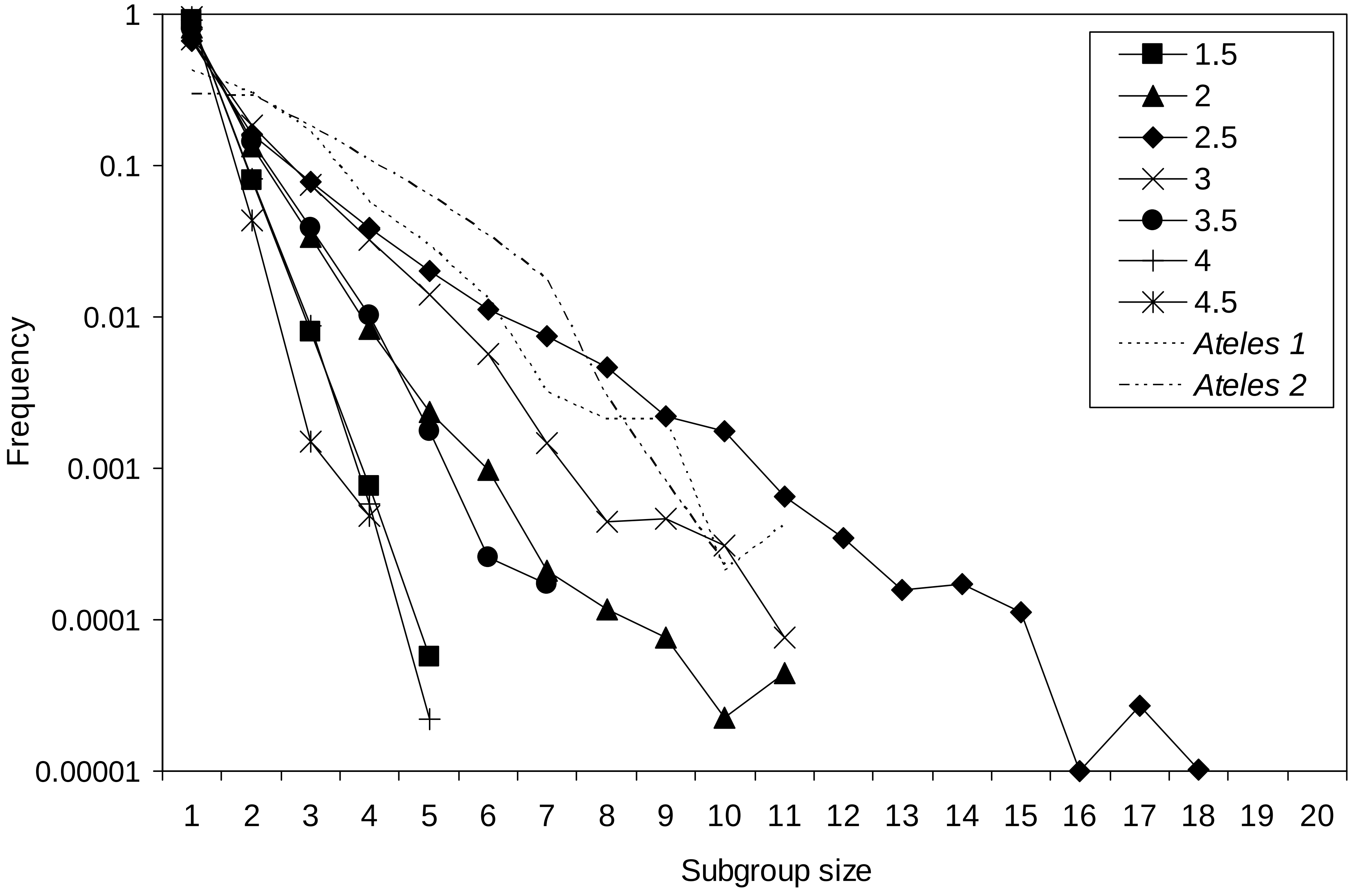

Frequency
1
0.1
0.01
0.001
0.0001
0.00001
1 2 3 4 5 6 7 8 9 10 11 12 13 14 15 16 17 18 19 20
Subgroup size
1.5
2
2.5
3
3.5
4
4.5
*Ateles 1*
*Ateles 2*

Figure 2b

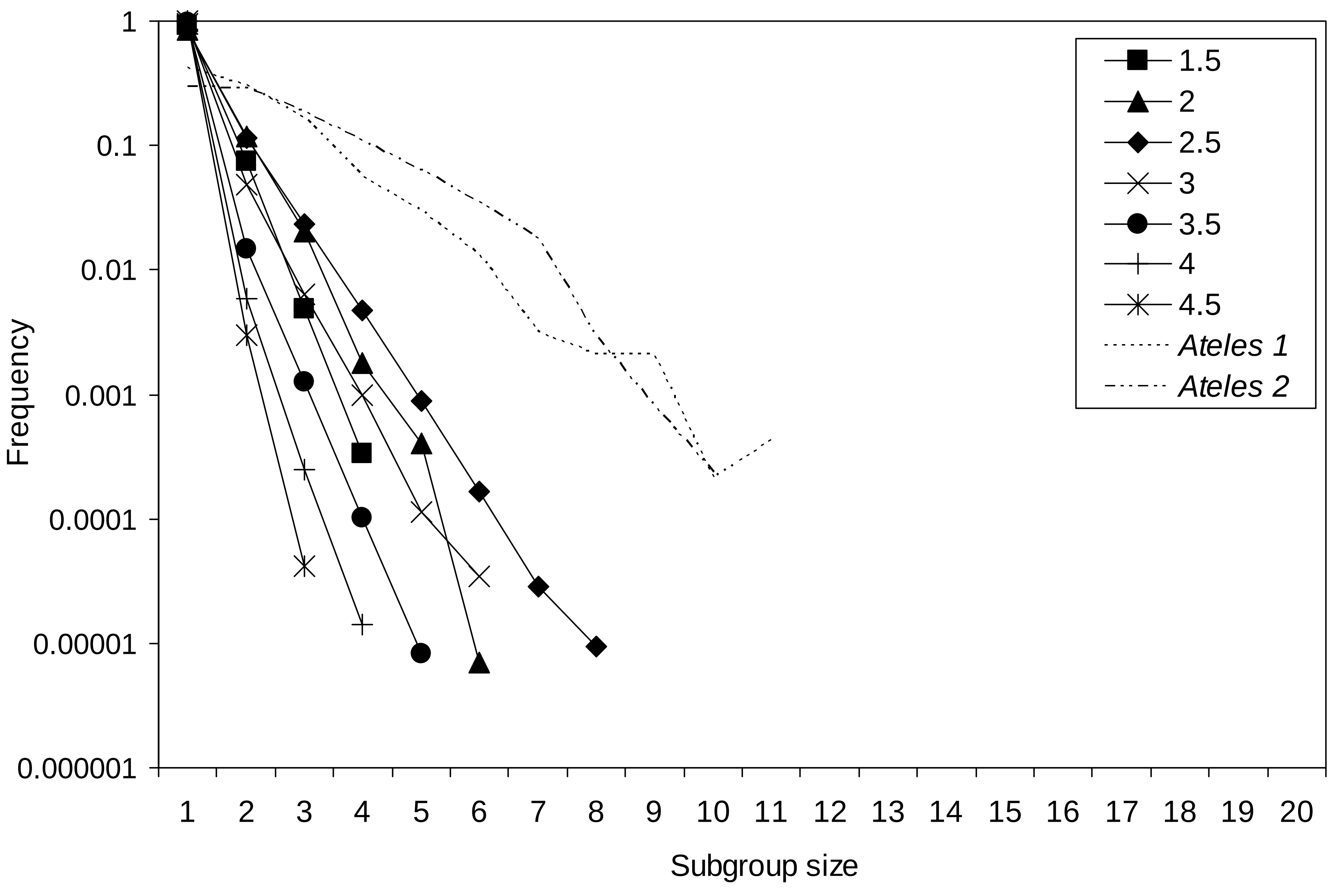

Frequency
1
0.1
0.01
0.001
0.0001
0.00001
0.000001
1 2 3 4 5 6 7 8 9 10 11 12 13 14 15 16 17 18 19 20
Subgroup size
1.5
2
2.5
3
3.5
4
4.5
Ateles 1
Ateles 2

Figure 2c

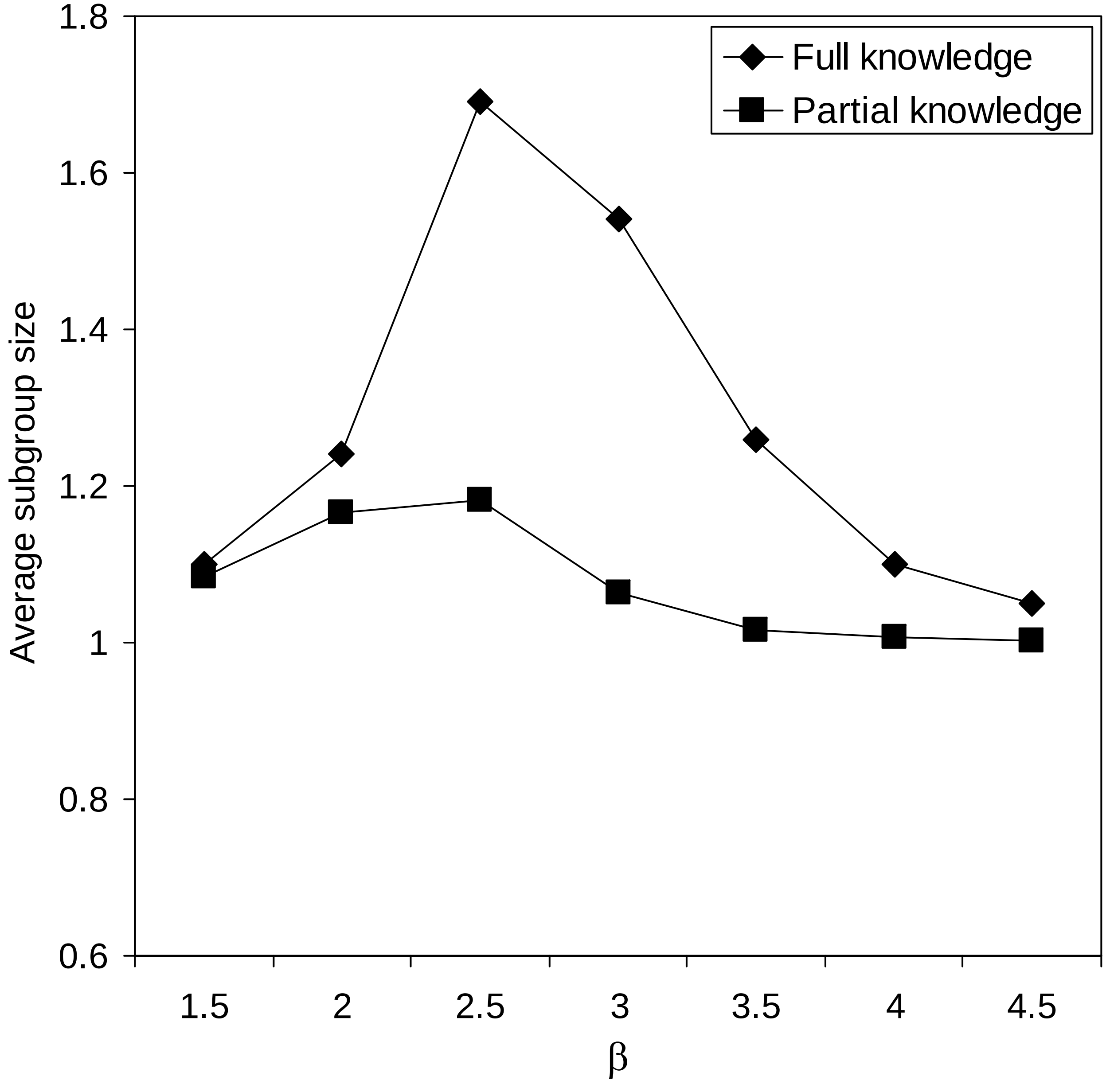

Figure 3a

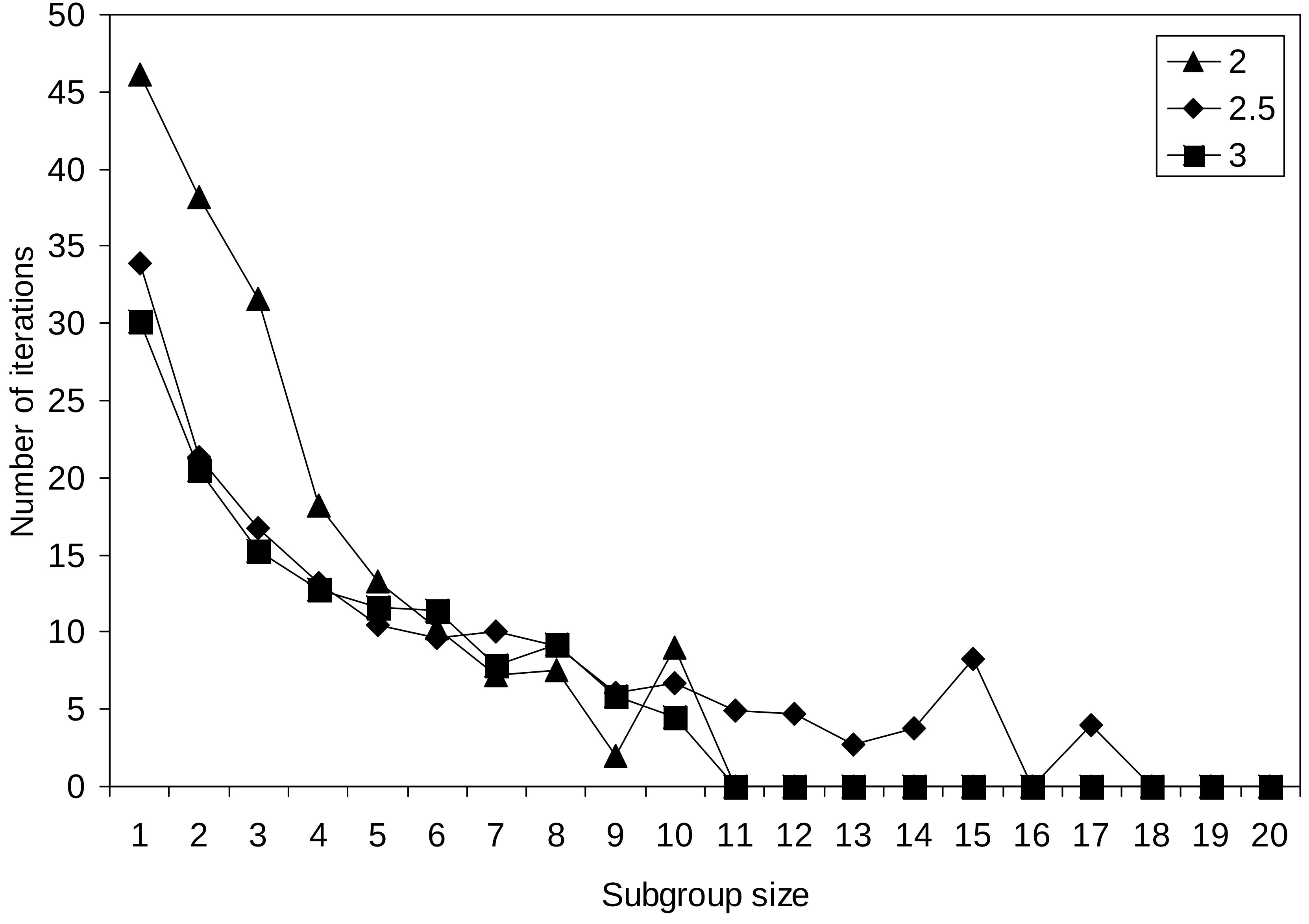

50
45
40
35
30
25
20
15
10
5
0
Number of iterations
1 2 3 4 5 6 7 8 9 10 11 12 13 14 15 16 17 18 19 20
Subgroup size
2
2.5
3

Figure 3b

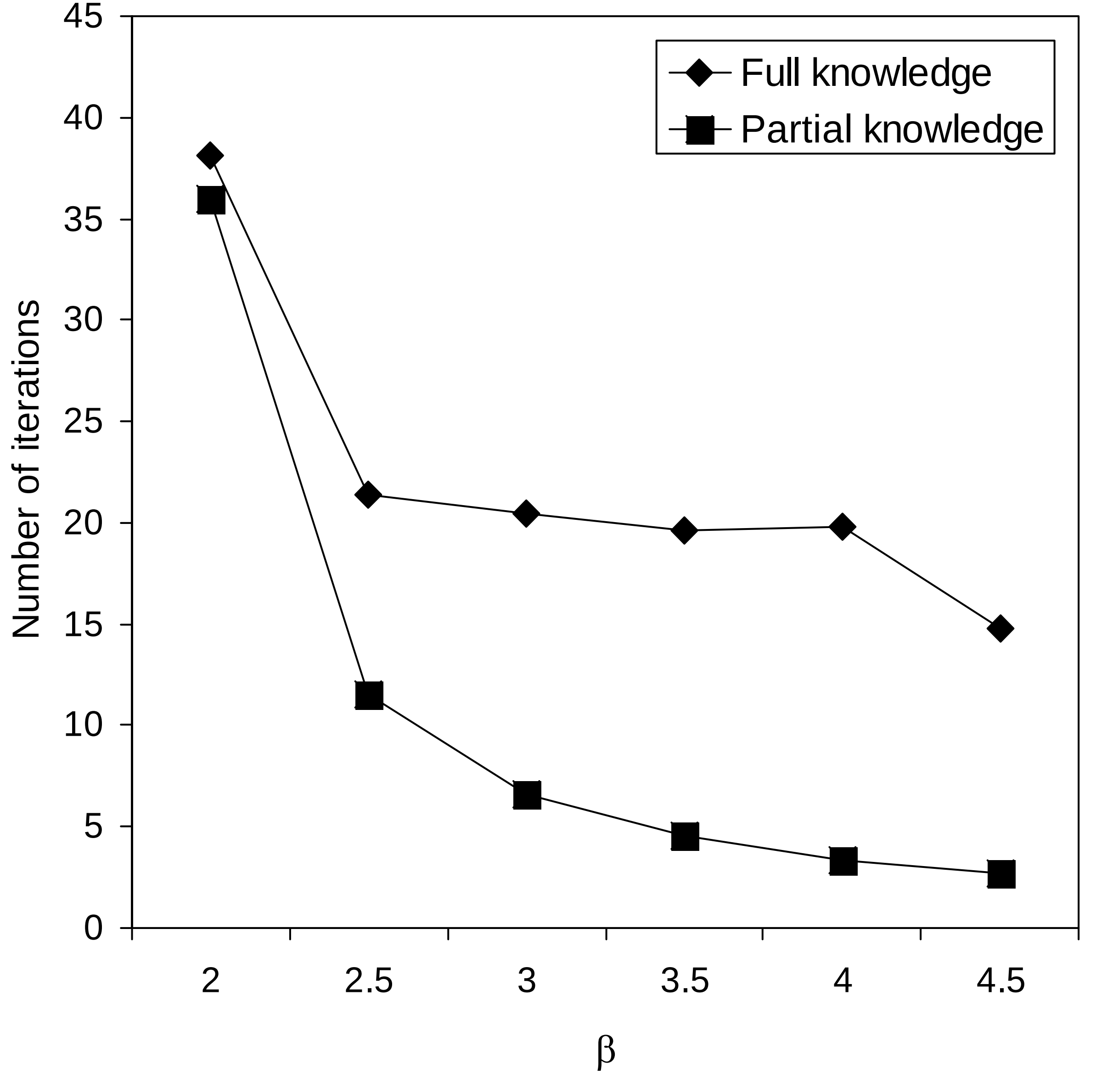

Figure 4a

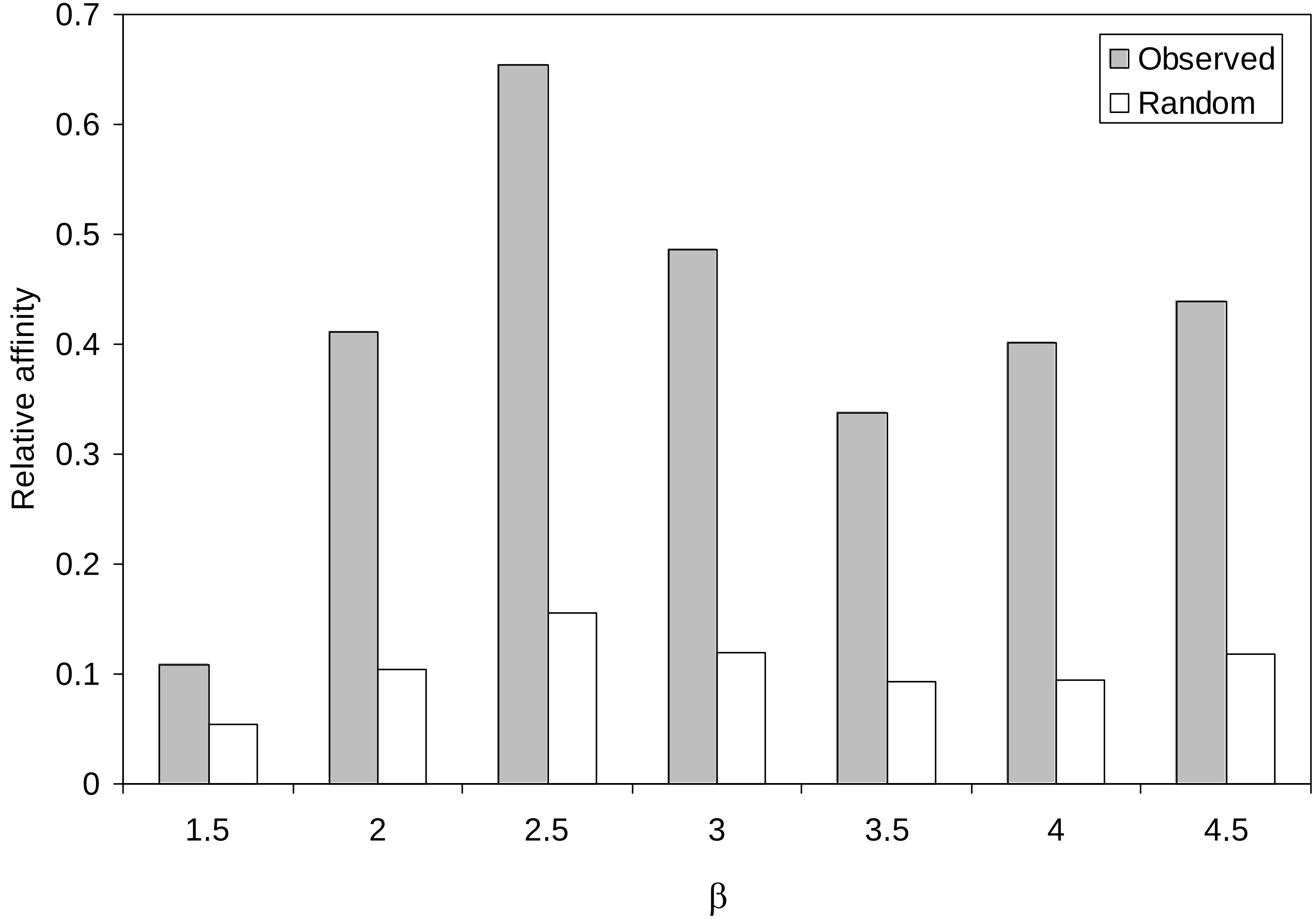

Figure 4b

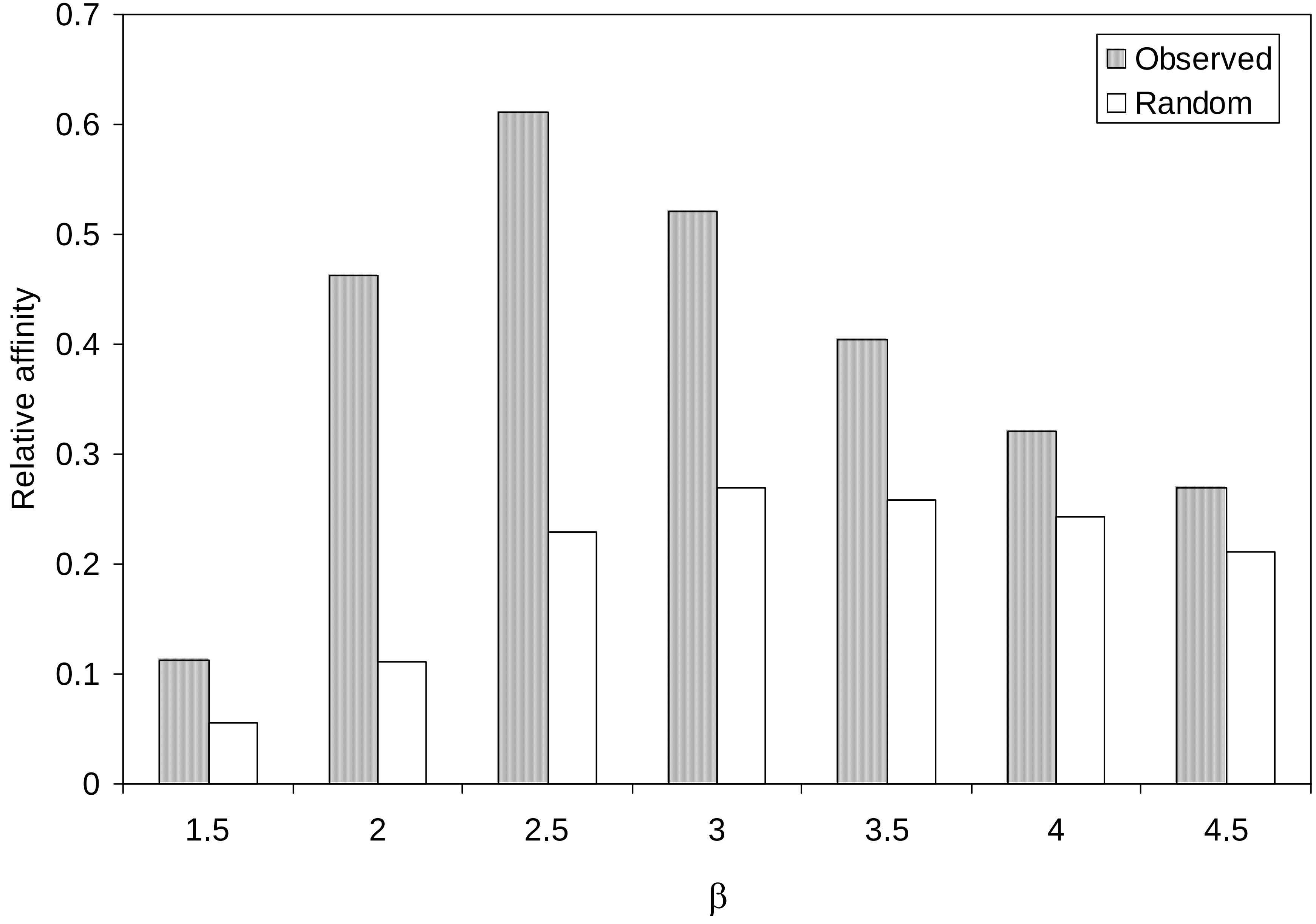

Figure 5a

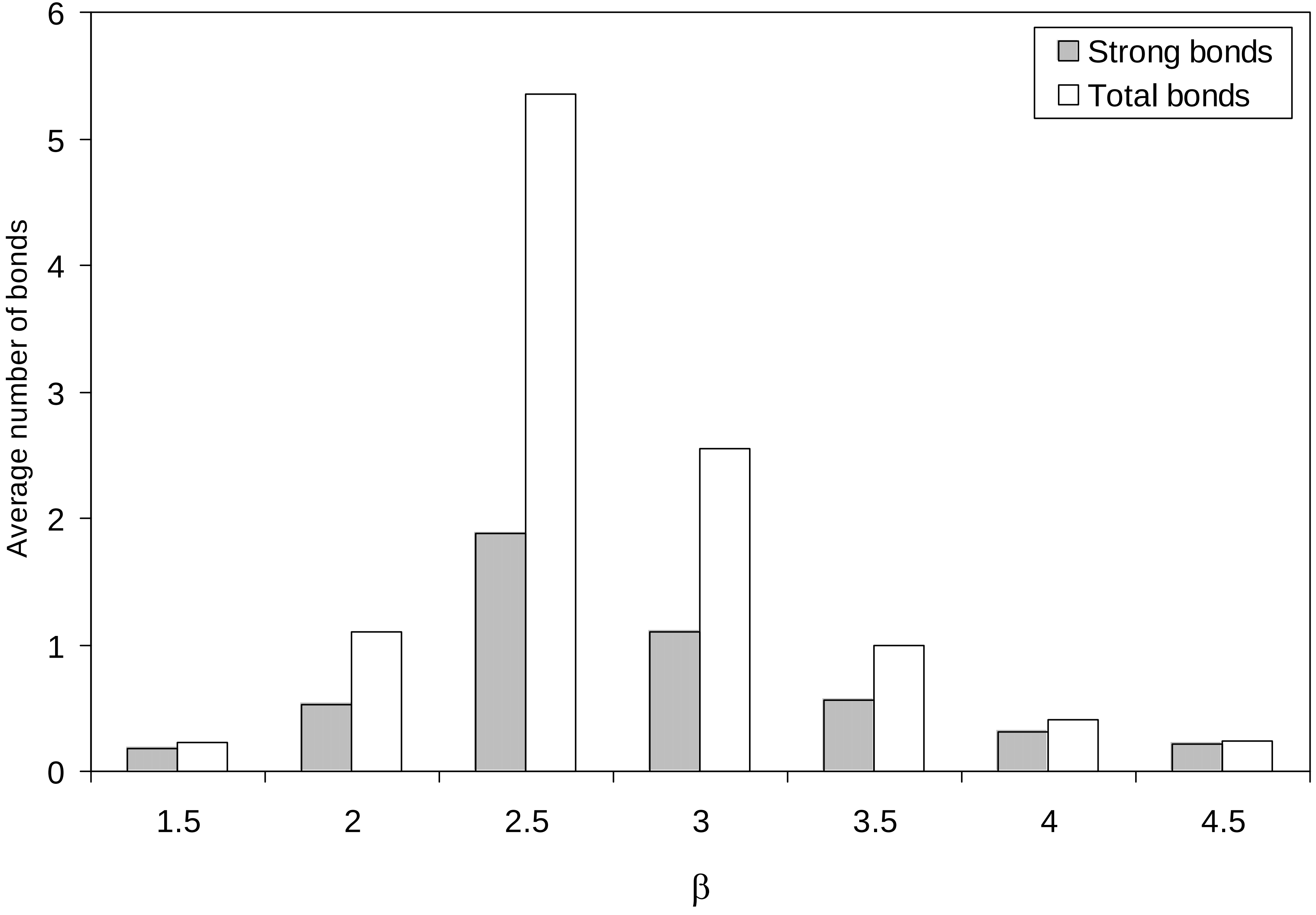

Figure 5b

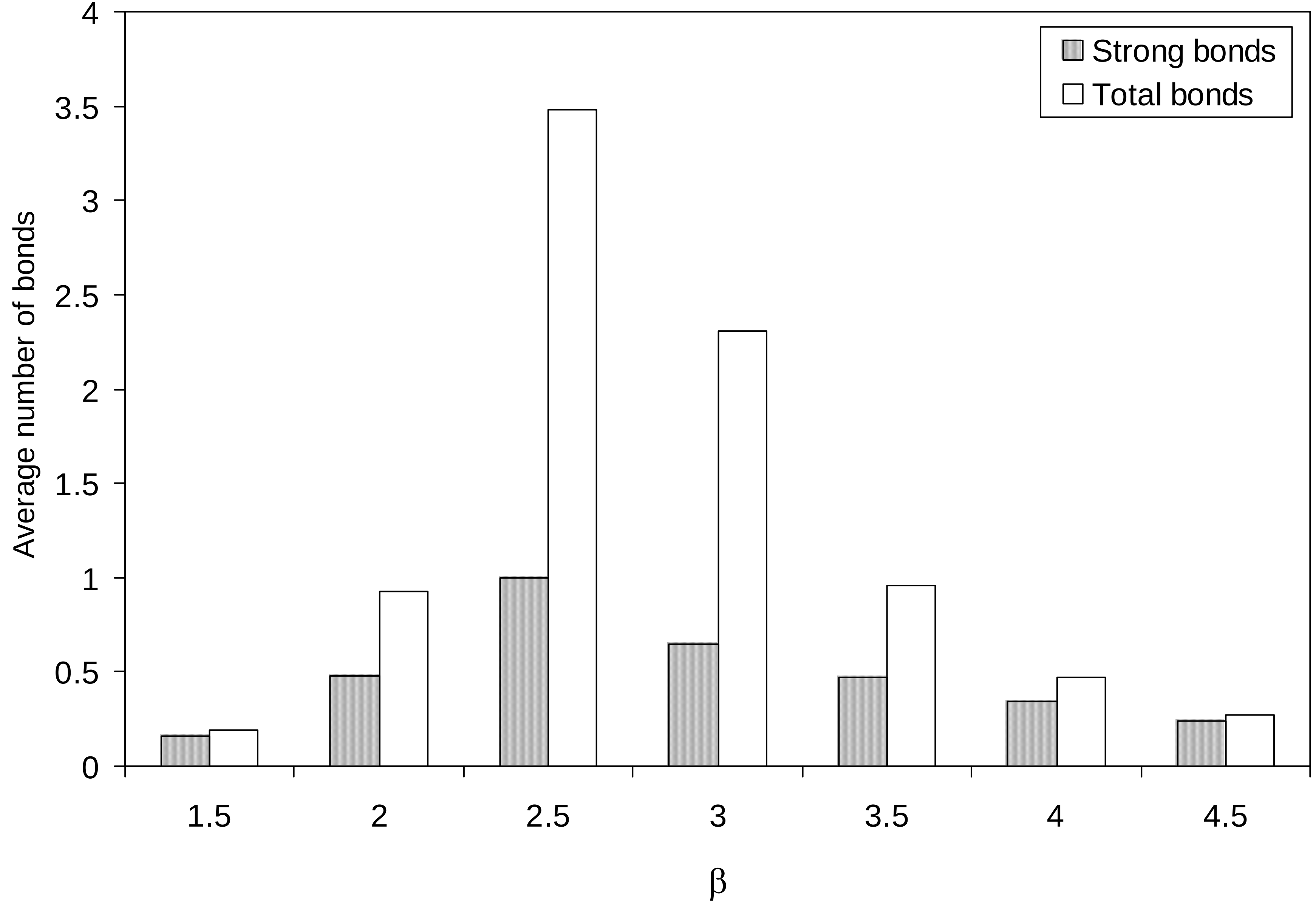

Figure 5c

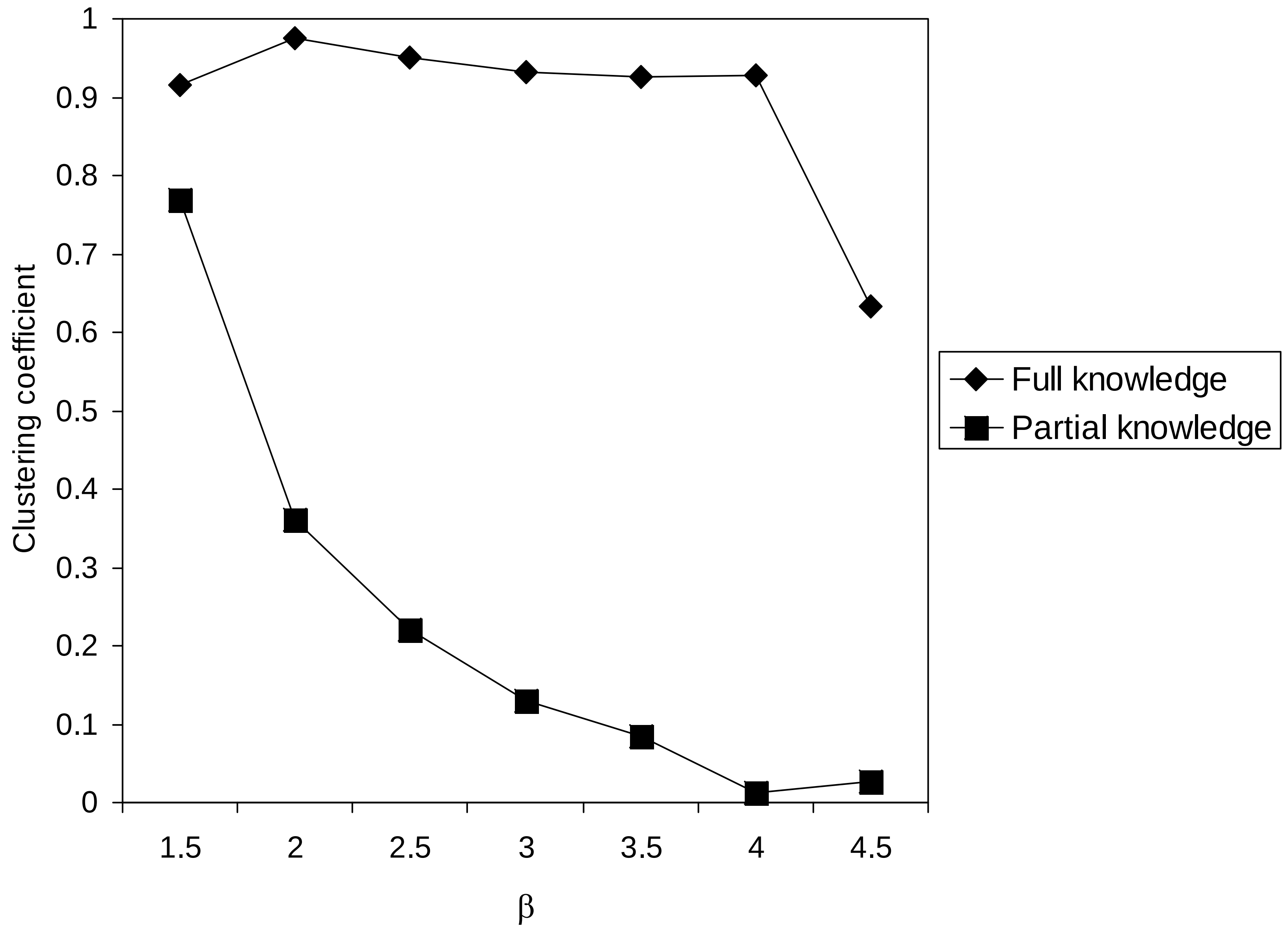

Figure 5d

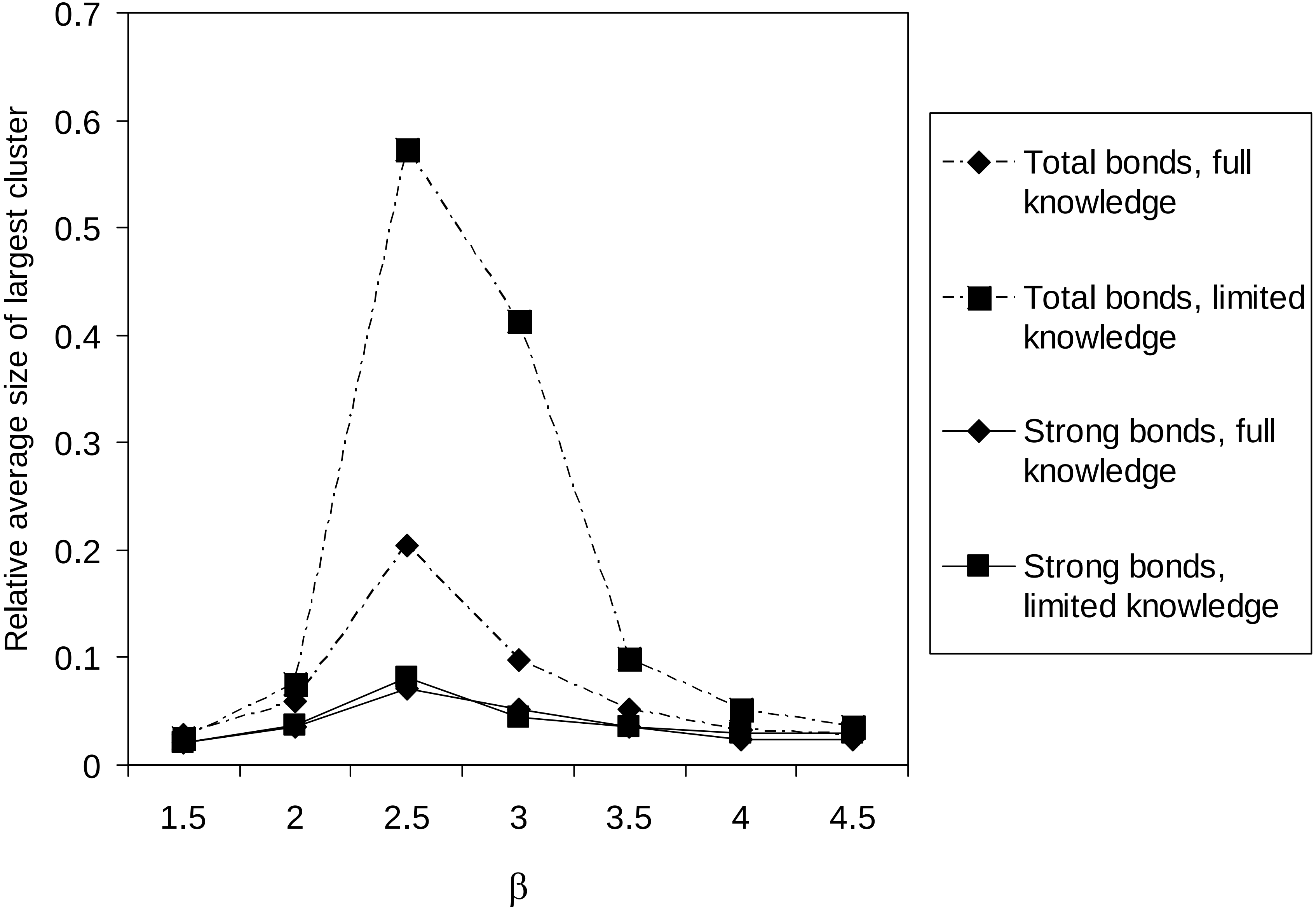
Relative average size of largest cluster
0.7
0.6
0.5
0.4
0.3
0.2
0.1
0
1.5
2
2.5
3
3.5
4
4.5
β
Total bonds, full knowledge
Total bonds, limited knowledge
Strong bonds, full knowledge
Strong bonds, limited knowledge

Figure 6

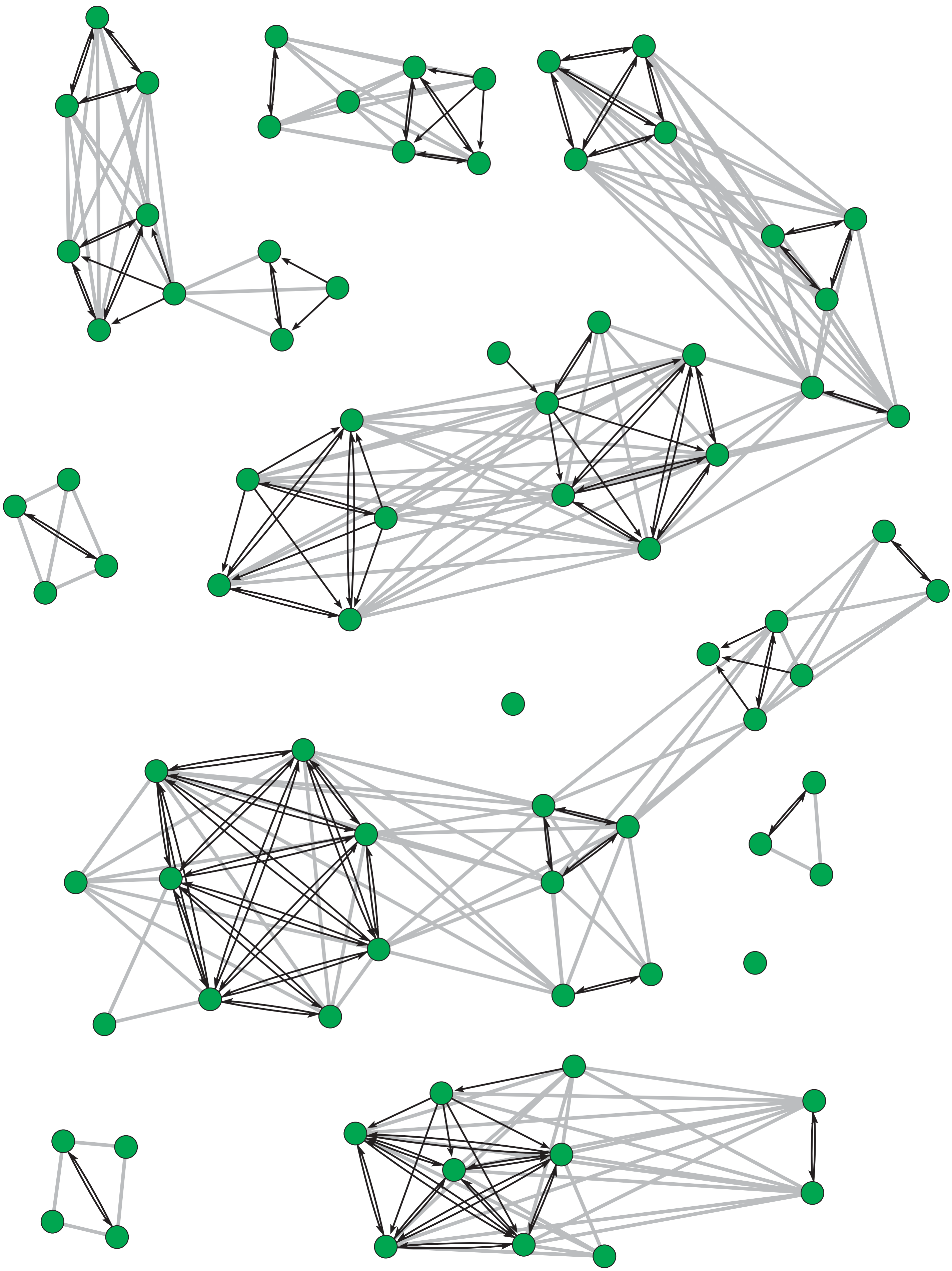

Table 1

| Extent of knowledge | Variation in tree size<br>**Large**<br>$\beta = 1.5 – 2$ | **Intermediate**<br>$\beta = 2.5 – 3$ | **Small**<br>$\beta = 3.5\text{-}4.5$ |
|---|---|---|---|
| Full | Small/medium subgroups<br>Long lasting (“frozen”)<br>Even relative affinity<br>Few strong bonds<br>Very cliquish<br>Non-percolating network | Large subgroups<br>Medium duration<br>Skewed relative affinity<br>Many strong bonds<br>Very cliquish<br>Percolating network | Small subgroups<br>Medium-short duration<br>Even relative affinity<br>Few strong bonds<br>Moderately cliquish<br>Non-percolating network |
| Partial | Very small subgroups<br>Long lasting (“frozen”)<br>Even relative affinity<br>Few strong bonds<br>Cliquish<br>Non-percolating network | Small subgroups<br>Medium-short duration<br>Skewed relative affinity<br>Few strong bonds<br>Moderately cliquish<br>Percolating network | Very small subgroups<br>Very short duration<br>Even relative affinity<br>Few strong bonds<br>Not cliquish<br>Non-percolating network |